\documentclass{aa}  
\usepackage{graphicx}
\usepackage{txfonts}
\usepackage{graphicx}
\usepackage{grffile}
\usepackage{txfonts}
\usepackage{amsmath}
\usepackage{epstopdf}
\usepackage{url}

\begin{document} 
\title{Dust properties of the cometary globule Barnard 207 (LDN 1489)\thanks{The reduced images (FITS file) in UV and optical bands are only available at the CDS via anonymous ftp to cdsarc.u-strasbg.fr (130.79.128.5) or via http://cdsarc.u-strasbg.fr/viz-bin/qcat?J/A+A/xxx/yy}}
\author{Aditya Togi\inst{1, 2}, A. N.~Witt\inst{1}, Demi St.John\inst{3, 4}}
\institute{Ritter Astrophysical Research centre, The University of Toledo, 2825 West Bancroft Street, Toledo, OH 43606, USA\\
\email{aditya.togi@utoledo.edu}
\and
Department of Physics and Astronomy, The University of Texas at San Antonio, One UTSA Circle, San Antonio, TX 78249, USA
\and
Institute of Engineering, Murray State University, Murray, KY 42071, USA
\and
Department of Physics, Montana State University, P. O. Box 173840, Bozeman, MT 59717, USA
}

\date{Received 1 August 2016 / Accepted 4 July 2017}

\abstract 
{Barnard 207 (B207, LDN 1489, LBN 777), also known as the Vulture Head nebula, is a cometary globule in the Taurus-Auriga-Perseus molecular cloud region. B207 is known to host a Class I protostar, IRAS 04016+2610, located at a projected distance of $\sim$8,400 au from the dense core centre. Using imaging and photometry over a wide wavelength range, from UV to sub-mm, we study the physical properties of B207 and the dust grains contained within. The core density, temperature, and mass are typical of other globules found in the Milky Way interstellar medium (ISM). The increase in the dust albedo with increasing optical wavelengths, along with the detection of coreshine in the near infrared, indicates the presence of larger dust grains in B207. The measured optical, near-, mid- and far-infrared intensities are in agreement with the CMM+AMM and CMM+AMMI dust grain type of The Heterogeneous dust Evolution Model for Interstellar Solids (THEMIS), suggesting mantle formation on the dust grains  throughout the globule. We investigate the possibility of turbulence being responsible for diffusing dust grains from the central core to external outer layers of B207. However, in situ formation of large dust grains cannot be excluded.}

\keywords{ISM: dust, emission, extinction -- ISM: molecules -- ISM: general -- ISM: individual obects: B207, LDN1489}
\authorrunning{Togi, Witt \& St.John}
\titlerunning{Dust properties in B207}
\maketitle

\section{Introduction}
Dust is one of the most important components of the galactic interstellar medium (ISM). Dust grains are the building blocks for planets to be formed in the later stages of the star formation process. In the diffuse interstellar medium the dust grain sizes follow the Mathis Rumpl \& Nodrsieck (MRN) distribution, a power law distribution dn/da $\propto$ a$^{-3.5}$ \citep{Mathis77}, where dn is the number of dust grains between sizes in the range a--a+da. According to the MRN distribution, grain sizes are in the range 0.005--0.25 $\mu$m. In denser interstellar environments, for example molecular clouds and globules, grain growth may occur as a result of the conglomeration of smaller grains and ice mantle formation. Such grains exhibit a less steep extinction curve (R$_{V}$ = 5.5) and increased efficiency of scattering at larger wavelengths. Recent studies have indicated that grains in low-mass molecular cores scatter mid-infrared (3.6--4.5 $\mu$m) light, termed coreshine \citep{Steinacker10, Pagani10} to differentiate it from cloudshine, which is observed at optical wavelengths from the outer, transparent parts of cores \citep{Foster06, Padoan06, Juvela06}. Coreshine has been interpreted as an effect in which starlight in the 3--5 $\mu$m wavelength range is efficiently scattered by larger than normal ISM grains. The dense cores of molecular clouds frequently exhibit this behaviour. The dense cores are sufficiently optically thick to prevent UV photons from penetrating and exciting polycyclic aromatic hydrocarbon molecules (PAHs), thus ruling out the possibility of 3.3 $\mu$m PAH emission contributing to the 3--5 $\mu$m coreshine. \citet{Pagani10} investigated a sample of 110 cores from which 95 cores were detectable in the Spitzer-IRAC band at 3.6 $\mu$m and about 50 clearly had signatures of coreshine. The emission at 3.6 $\mu$m is spatially coincident with the densest regions of the cores. Recent studies by \citet{Jones12a, Jones12b, Jones12c, Jones13, Kohler14, Ysard15} have suggested that the cloudshine and coreshine can be explained by dust growth through the formation of aliphatic-rich amorphous hydrocarbon mantles. They used The Heterogeneous dust Evolution Model for Interstellar Solids (THEMIS) and a radiative transfer model, to study the effects of mantle formation with minimal coagulation to explain the origin of cloudshine and coreshine.

Here, we study the dust properties from optical to sub-mm wavelengths and the observed coreshine phenomenon in one particular core, Barnard 207 (B207, LDN 1489). B207 is located at a distance of 140 pc in the Taurus-Auriga-Perseus molecular cloud region \citep{Loinard05, Kenyon94}. B207 is a high Galactic latitude Bok globule, with equatorial coordinates (J2000.0) RA = 61.216$^{\circ}$ Dec = 26.318$^{\circ}$ (galactic coordinates: $\ell$ = 168.08$^{\circ}$, b = -19.15$^{\circ}$). A 1.6 M$_{\odot}$ protostellar source, IRAS 04016+2610, is currently forming at the west side of B207's core \citep{Yen14}. This protostar is located at a projected distance of about 8,400 au from the centre of the present-day dense core of B207. \citet{Benson98}, using high spatial and spectral resolution of N$_{2}$H$^{+}$ and C$_{3}$H$_{2}$, conclude that the line shapes for B207 are consistent with models of cloud cores undergoing gravitational collapse, but observations indicate the core to be as yet starless, suggesting that the formation of a new star is in its earliest stages.

The paper is structured as follows. In Sect. 2 we discuss our optical observations and the archival infrared data. In Sect. 3 we describe our analysis. Section 4 discusses our results and finally we summarize our findings in Sect. 5.

\section{Observations \& archival data}
The objective of the observations is to measure the surface brightness of the scattered light, and to study the general dust properties of the cloud in different optical and infrared bands. The far-infrared and sub-mm data will be used to determine physical properties such as densities and temperatures in different parts of B207 to better associate grain properties with the environmental conditions.We will be looking for evidence of grain growth.

\subsection{Discovery Channel Telescope (DCT) observations}
B207 was imaged in several optical bands using the Large Monolithic Imager \citep[LMI;][]{Massey13} of the 4.3-metre Discovery Channel Telescope \citep[DCT;][]{Degroff14}, operated by Lowell Observatory, on 2013 February 5-6 and 2013 December 5. LMI exposures were taken using its charged-coupled device (e2v CCD231), having 6144 $\times$ 6160 pixels, with 2 $\times$ 2 binning. Imaging was performed with the Johnson UBV and Kron-Cousins R and I filters, at effective wavelengths 366, 428, 537, 633, and 806 nm, respectively (these calculated wavelengths are the arithmetic means of the two wavelengths at which the transmission is half of the maximum\footnote{\url{www2.lowell.edu/rsch/LMI/specs.html}} in each filter). The LMI field-of-view (FOV) was 12.5$\arcmin$ $\times$ 12.5$\arcmin$, large enough to cover most of the dense parts of the globule as well as adjacent sky regions. The plate scale was 0.24$\arcsec$/pixel for 2 $\times$ 2 binning. 

The globule was observed with total exposure times of 180, 240, 105, and 75 minutes in U, B, V, and R bands, respectively, consisting of sub-exposures of 15--20 minutes. In the I band, two exposures were taken, each of nine minutes duration. The observable characteristics of the globule are represented in Fig. 1, showing a three-colour stacked image in which blue, green, and red colours represent the V, R, and I bands, respectively. The basic data reductions such as flat fielding, bias correction, and sky subtraction were performed using the Image Reduction and Analysis Facility (IRAF). The effects of cosmic rays in the images were removed using the L.A. Cosmic package \citep{Dokkum01} within IRAF. The photometric calibration was derived using a standard star, GD71, from the list of \citet{Landolt92}, located close to the globule's location. Using the LMI-DCT manual, the photometric scale was fixed and the zero points were determined. The magnitudes of other stars in the field were determined and found to be consistent, within 10$\%$ uncertainty in the observed bands, when compared to entries in SIMBAD.

\begin{figure}
\includegraphics[width=0.5\textwidth]{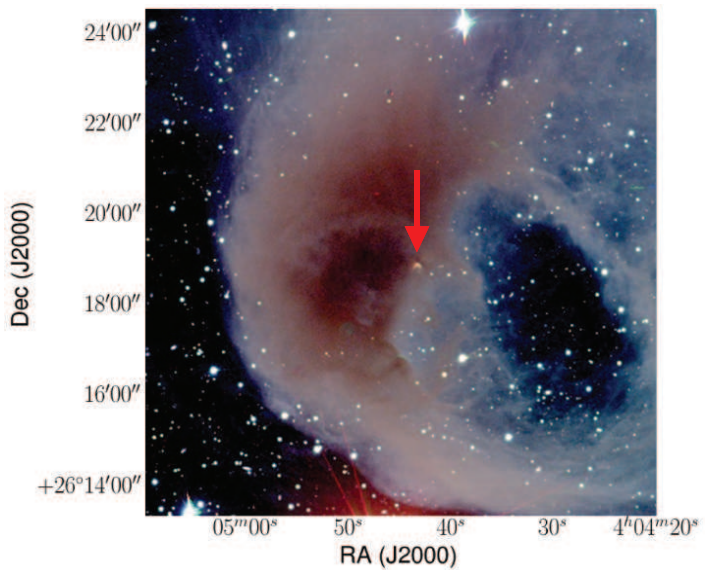}
\caption{Three-colour stacked image of B207. The blue, green, and red colours represent the V, R, and I bands, respectively. The protostar, IRAS 04016+2610, at  R.A. = $4^{h}4^{m}43.071^{s}$ , Dec = $+26^{\circ}18\arcmin56.39\arcsec$ is forming at a distance of about 8,400 au west of the core, marked by the red arrow. A sharp boundary is seen towards the outer rim of the cloud to the east. A less dense region of material, appearing as a hole, is located towards the immediate west of the protostar.}
\label{fig:b207}
\end{figure}

\subsection{Archival 2MASS, WISE, and Herschel data}
We employed archival Two Micron All Sky Survey \citep[2MASS;][]{Skrutskie06} J, H, K$_{S}$ band images along with Wide-field Infrared Survey Explorer \citep[WISE;][]{Wright10} W1 and W2 image data at 3.4 and 4.6 $\mu$m to investigate the presence of coreshine in the inner regions of the B207 globule. The 2MASS archival processed images have a pixel scale 1.0$\arcsec/pixel$. The WISE band images were 10$\arcmin$ $\times$ 10$\arcmin$ wide. The pixel scale was 1.375$\arcsec/pixel$ in the W1 and W2 bands. 
To determine the spectral energy distribution (SED) of the core and the rim of the globule, we used archival Herschel Space Observatory's  Photoconducting Array Camera and Spectrometer \citep[PACS;][]{Poglitsch10} -100 and 160 $\mu$m images with Spectral and Photometric Imaging Receiver \citep[SPIRE;][]{Griffin10} - 250, 350, and 500 $\mu$m maps observed in the \emph{Herschel- Key Program Guaranteed Time (KPGT)} (PI Ph. Andre). The full width half maximum (FWHM) point spread functions (PSFs) for WISE1, WISE2, PACS 100, 160, SPIRE 250, 350, and 500 $\mu$m are 5.84$\arcsec$, 6.48$\arcsec$, 7.70$\arcsec$, 12.8$\arcsec$, 17.6$\arcsec$, 23.9$\arcsec$, and 35.2$\arcsec$, respectively.

\section{Analysis}
The mass, temperature, and density distribution, along with other dust and gas properties, provide hints to the past, present, and future evolutionary fate of a globule. They also define the environmental conditions under which the dust grains exist and evolve. In this section, we provide a description of our methods for determining the physical properties of B207.

B207 shows many interesting features, as seen in Fig. \ref{fig:b207}. First, a protostellar object (IRAS 04016+2610) is seen near the centre of the image, at a projected offset of $\sim$ 8,400 au  from the dark core of B207. Second, a nearly transparent region of gas and dust is seen $\sim 30,000$ au towards the west of the newly forming star. Third, a sharp edge is seen towards the eastern side of the cloud. Fourth, a small dark core in B207 is seen as a dark region to the east of IRAS 04016+2610. The bright rim surrounding the core is typical for high-latitude globules and is interpreted as resulting from scattering of the galactic radiation field in the optically thin outer portions of the globule \citep{Fitzgerald76}.

\subsection{Physical parameters of B207: SED fit}
\begin{figure}
\includegraphics[width=0.47\textwidth]{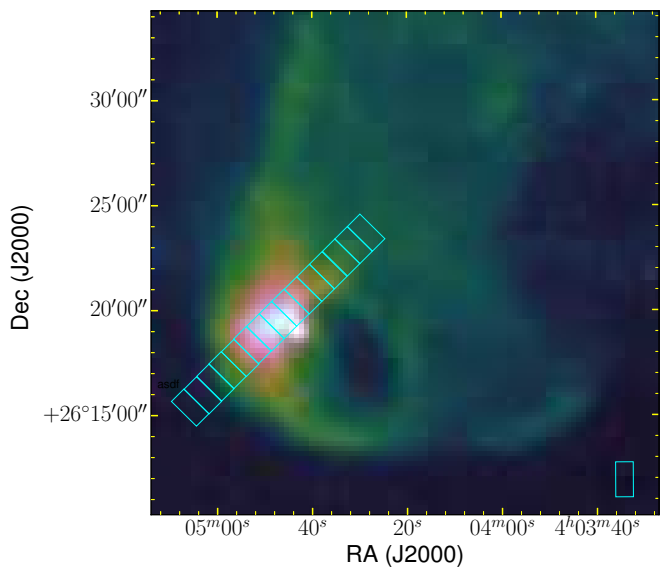}
\caption{Overlay of our rectangular cut region on the core of B207 on a SPIRE-Herschel dust emission map at 500 $\mu$m. The rectangular box towards the lower right in the image measures the background intensity.}
\label{fig:b207_500}
\end{figure}

\begin{figure}
\includegraphics[width=0.48\textwidth]{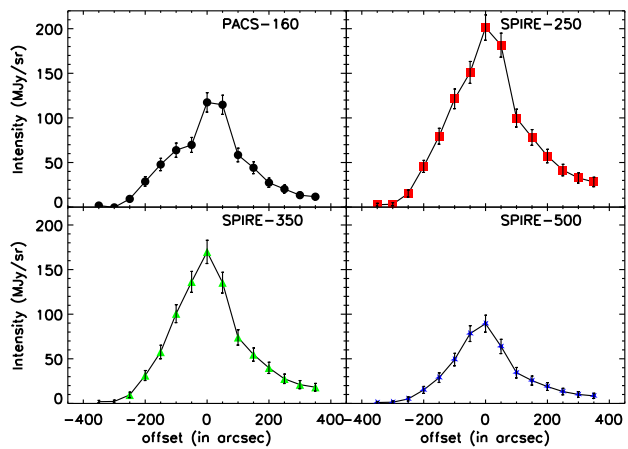}
\caption{Measured background subtracted integrated FIR intensities for each rectangular region as a function of core distance. The zeroth position is the centre of the central rectangular box with equatorial coordinates RA = $4^{h}4^{m}47^{s}$ , Dec = $+26^{\circ}19\arcmin32\arcsec$.}
\label{fig:firflux}
\end{figure}

We used the PACS 160 $\mu$m and SPIRE 250, 350, and 500 $\mu$m dust emission maps obtained from the $\emph{Herschel Space Observatory}$ to model the line-of-sight average dust temperature and column density of B207. A diagonal cut, shown in Fig. \ref{fig:b207_500}, parallel to the orientation of the elongated core, was made to estimate the average individual background-subtracted intensities over the strip of 15 boxes, each of size 50$\arcsec\times$100$\arcsec$. These measured values of FIR intensities at 160, 250, 350, and 500 $\mu$m are plotted in Fig. \ref{fig:firflux}. The core centre is at the  zeroth spatial position in the plot with negative and positive offsets signifying the direction from southeast to northwest.

Spectral energy distributions (SEDs) were constructed and single temperature modified black-body curves were used to fit the SEDs for each rectangular box region . We used a single dust temperature black-body model $S_{\nu} = \Omega B(\nu,T_{d})(1-e^{-\tau(\nu)})$, where $\Omega$, $B(\nu,T_{d})$, and $\tau(\nu)$ are solid angle of the emitting element, Planck function at dust temperature T$_{d}$, and optical depth at frequency $\nu$, respectively. We used a dust opacity model, $\kappa_{\nu} \propto \nu^{\beta}$ with $\beta = 1.8$ \citep{Planck11} to fit the SED using the MPFIT programme \citep{Markwardt09}. 

We used the \emph{Herschel-SPIRE} 500 $\mu$m map to estimate the column density. We calculated B(500$\mu$m, T$_{d})$ from dust temperatures estimated from the SED fit for each region to calculate the optical depth $\tau_{500}$ = $\frac{I_{500}}{B_{500 \mu m}}$. The column density, N(H), can be obtained from
\begin{equation}
N(H) = \frac{M_{H}}{M_{d}} \times \frac{\tau_{500}}{\kappa_{500}\times m_{H}},
\end{equation}
where $\frac{M_{H}}{M_{d}} = 110$ is the adopted hydrogen gas-to-dust mass ratio in the solar neighbourhood \citep{Sodroski97} and $\kappa_{500}$ is the dust opacity at 500 $\mu$m. We used the value for $\kappa_{500} = 2.85$ cm$^{2}$ g$^{-1}$ from \citet{Kohler15}, average for AMM and AMMI type dust grains. 
\begin{figure}
\includegraphics[width=0.48\textwidth]{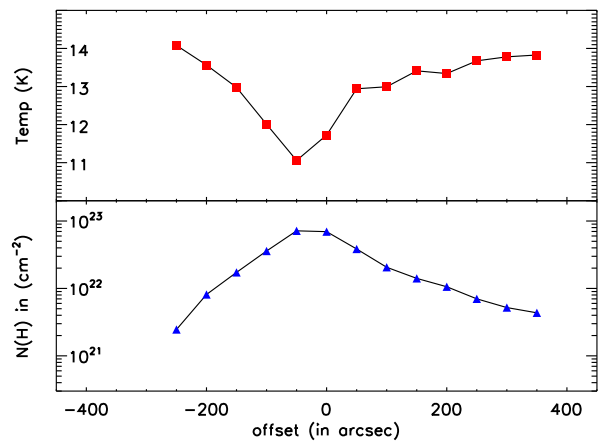}
\caption{Temperature profile of B207 obtained by fitting the SED at the FIR wavelengths. The temperature decreases to about 11 K in the core compared to the surrounding medium of 14 K. We used these estimated temperatures in Planck's function to calculate optical depths at 500 $\mu$m to estimate column densities. The column density, N(H), at the core is about 7.2$\times$10$^{22}$ cm$^{-2}$, causing an extinction of about A$_{V}$ = 48, assuming R$_{V}$ = 5.5 for molecular clouds, and assuming the aggregate (AMM and AMMI type) dust grain opacity from \citet{Kohler15}.}
\label{fig:tempden}
\end{figure}

The spatial width of each box is 50$\arcsec\times$100 $\arcsec$ = 0.034$\times$0.068 pc$^{2}$ for an assumed distance of 140 pc. After estimating the column density, we calculated the mass of the gas column in each region. The FIR intensities, temperature, column density, and mass of the gas of each  region are listed in Table \ref{table:firint}. Summing up the mass estimates, the total gas mass is about 5.6 M$_{\odot}$ (not including heavy element correction). Figure \ref{fig:tempden} shows the temperature and the column density profile for B207 over our selected regions. The temperature at the core is about 11 K and the peak column density, N(H), is 7.2$\times$10$^{22}$ cm$^{-2}$. This column density is equivalent to an extinction of A$_{V}$ = 48, assuming R$_{V}$ = 5.5 for renormalized case A from \cite{WD01, Draine03}. When considering an alternate dust mass opacity  $\kappa_{500}$ = 5.04 cm$^{2}$/g at 500 $\mu$m for dust grains with thin ice mantles at central density 10$^{6}$ cm$^{-3}$ \citep{Ossenkopf94}, the column density and mass in each region was reduced by a factor $\frac{5.04}{2.85}$=1.77 and the total mass reduces from 5.6 to 3.2 M$_{\odot}$. This illustrates the approximate nature of these estimates.

\begin{table*}
\centering
\begin{minipage}{124mm}
\caption{Measured FIR intensities of B207 and derived physical properties}
\begin{tabular}{cccccccccc}
\hline
Region & 160$\mu$m & 250$\mu$m & 350$\mu$m & 500$\mu$m & T & N(H) & M(H) & T$_{Abel}$ & N$_{Abel}$(H) \\
 & MJy/sr & MJy/sr & MJy/sr & MJy/sr & K & 10$^{21}$ cm$^{-2}$ & M$_{\odot}$ & K & 10$^{21}$ cm$^{-2}$ \\
\hline
1(south)& 1.75 & 2.66 & 1.97 & 1.01 & -- & -- & --\\
2       & 0.30 & 3.28 & 2.21 & 1.36 & -- & -- & --\\
3       & 9.21 & 15.3 & 9.54 & 5.00 & 14.1   & 2.46  &  0.04 & 11.6 & 4.01\\
4       & 28.7 & 45.6 & 31.3 & 15.1 & 13.6   & 8.13  &  0.15 & 11.1 & 13.7\\
5       & 47.8 & 79.5 & 57.6 & 28.9 & 13.0   & 17.3  &  0.32 & 10.6 & 29.9\\
6       & 63.8 & 121  & 101  & 49.1 & 12.0   & 35.9  &  0.66 & 10.1 & 58.5\\
7       & 69.8 & 151  & 136  & 78.2 & 11.1   & 71.6  &  1.31 & 9.40 & 117\\
8(core) & 117  & 201  & 170  & 89.4 & 11.7   & 69.6  &  1.27 & 10.1 & 107\\
9       & 115  & 182  & 135  & 63.9 & 12.9   & 38.5  &  0.70 & 10.6 & 66.0\\
10      & 58.5 & 99.8 & 74.0 & 34.4 & 13.0   & 20.6  &  0.38 & 11.1 & 31.2\\
11      & 44.1 & 78.1 & 54.8 & 25.6 & 13.4   & 14.1  &  0.26 & 11.6 & 20.5\\
12      & 27.4 & 57.3 & 39.9 & 18.9 & 13.3   & 10.6  &  0.19 & 12.7 & 12.0\\
13      & 20.3 & 41.5 & 27.6 & 13.3 & 13.7   & 7.01  &  0.13 & 13.3 & 7.50\\
14      & 13.4 & 32.7 & 21.1 & 10.1 & 13.8   & 5.20  &  0.09 & -- & --\\
15(north) & 11.7 & 28.2 & 18.2 & 8.48 & 13.8 & 4.35  &  0.08 & -- & --\\
\hline
\end{tabular}
{Temperature, column density and mass for regions 1 and 2 were not estimated since they are outside of the globule in the diffuse sky region.\\
The column density, N(H), and mass, M(H), are calculated for each region using the AMM and AMMI dust mass opacity at 500 $\mu$m from \citet{Kohler15}.\\
The temperature, T$_{Abel}$, and column density, N$_{Abel}$(H), are the temperature and the corresponding column density, respectively,  estimated from the Abel inversion method.}
\label{table:firint}
\end{minipage}
\end{table*}

\subsection{Temperature and number density-Abel Inversion method}
Recently, the radial variation in dust temperature and density has been studied in a few starless cores. \citet{Roy14}, using the inverse Abel transform technique on FIR Herschel maps of B68 and L1689B, concluded that the actual core dust temperature is about 2 K lower compared to the temperature obtained by the line-of-sight averaged SED fitting, which we used in Sect. 3.1. Evaluating the core temperature using a SED fit gives an average temperature along the line of sight and does not take into account the temperature gradients within the sources. We adopted the inverse Abel measurement technique to study the variation in temperature along the radial direction towards the core of B207. We used \emph{Herschel-SPIRE} FIR maps at 250 and 350 $\mu$m for this analysis. We direct the reader to \cite{Roy14} for the mathematical details of the method.

\begin{figure}
\includegraphics[width=0.48\textwidth]{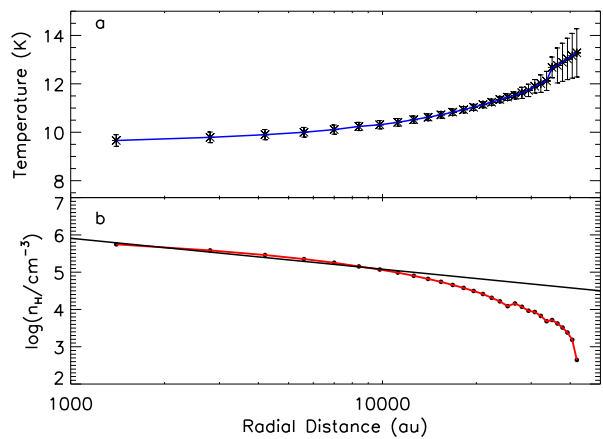}
\caption{a) Dust temperature profile of B207 obtained by applying the Abel inversion method to circularly-averaged intensity profiles observed with Herschel at 250$\mu$m and 350$\mu$m. The error bars represent the standard deviation of the mean in T$_{d}$(r) values obtained from independent profile reconstruction along 12 angular directions.
b) Density profile of B207 obtained by applying the Abel inversion method to circularly-averaged intensity profiles observed with Herschel at 250 $\mu$m.}
\label{fig:tdprofile} 
\end{figure}

We used convolved 250 and 350 $\mu$m \emph{Herschel-SPIRE} intensity values to evaluate the temperature profile of B207. The cloud was divided into 12 equal angular sector regions along with equal radial steps for each angular sector region. A radial temperature profile for each angular sector was computed. Figure \ref{fig:tdprofile}a represents the radial temperature profile of B207, averaged over all 12 angular sector regions. The error in the temperature value is computed using the standard deviation from 12 angular sector regions. However, because of the presence of the protostar, IRAS 04016+2610, and the transparent hole region, angular sector regions towards the western side of the core and at a distance greater than 10,000 au were not considered. The protostar is likely to heat gas and dust in its vicinity, causing deviations in the temperature profile of the globule. We will discuss the effects of the protostar on the core's physical properties in more detail in Sect. 4.5. 

The temperature profile, when extrapolated to the core, reaches a minimum of about 9.4$\pm$0.1 K. The temperature increases to 13.4$\pm$1.0 K in the outer regions of the cloud at a distance of about $\sim$ 40000 au, where it is more strongly heated by the interstellar radiation field (ISRF). The core temperature derived by the Abel inversion technique is 1.7 K less than that found from the SED fitting technique. Table \ref{table:temp} lists the core and rim temperatures for B207 derived from the SED fit and Abel inversion method. 

\begin{table}
\centering
\caption{B207 core and rim temperature.}
\begin{tabular}{ccc}
\hline
Region & T$_{SED}$(K) & T$_{Abel}$(K)\\
\hline
Core & 11.1 &  9.4$\pm$0.1\\
Rim  & 14.0 & 13.4$\pm$1.0\\
\hline
\end{tabular}
\label{table:temp}
\end{table}

Knowing the radial temperature distribution in B207 we calculated the radial density profile from the 250 $\mu$m map (using Equation 2 of \cite{Roy14}). We used the dust mass opacity at 250 $\mu$m, $\kappa_{250}$ = 11.45 cm$^{2}$/g for AMM and AMMI type dust grains \citep{Kohler15}. Assuming a hydrogen gas-to-dust mass ratio of 110 in the analysis, we adopted a gas mass opacity at 250 $\mu$m, $\kappa_{250,gas}$ = $\kappa_{250}$/110 = 0.10 cm$^{2}$/g. The density distribution was obtained again for each angular sector region. Figure \ref{fig:tdprofile}b represents the density profile for B207 averaged over all angular sector regions. The density distribution flattens out towards the core centre at about 5.6$\times$10$^{5}$ cm$^{-3}$ and is significantly different in slope from that in the outer regions of the cloud, at distances larger than 9800 au, shown by the deviation from the solid line in Figure \ref{fig:tdprofile}b. We list the column densities, N$_{Abel}$(H) estimated from the average calculated temperature for each box region using the Abel inversion method in the last column of Table \ref{table:firint}. The column densities at the core are about 1.6 times higher than those derived with the SED method.

\citet{Pagani15} have shown the existence of cold dust and gas (6--7 K) and concluded that the modelling technique used by \citet{Roy14} can potentially miss a large fraction of the mass (30--70$\%$) in cores. This cold dust is not identified solely by its emission as the warm dust dominates because of the non-linearity of the black-body function. A recent study by \cite{Hocuk17} parameterized the dust temperature with the UV radiation field. Adopting the value of $\chi_{uv} = 1.71$ \citep{Draine78, Draine11} and A$_{v}$ = 48 in the core of B207, the dust temperature at the core is estimated to be $\sim$7 K, about 2.4 K lower than the value estimated by the Abel inversion technique. Another study by \citet{Steinacker16a} has suggested that the Abel inversion technique yields core temperatures, which are still about 1-2 K higher than the actual core dust temperatures. If correct, this results in a further underestimation of core masses by a factor of approximately two. However, \citet{Steinacker16b} concluded the presence of a small mass fraction ($<$ 20$\%$) with temperatures $<$ 8 K in molecular cores. The use of a somewhat uncertain dust mass opacity at 500 $\mu$m and a constant gas to dust ratio ($M_{H}/M_{d}$) in our mass estimate analysis is subject to a random error of similar magnitude along with the systematic error of the Abel inversion technique. These studies indicate ongoing efforts in understanding core dust temperatures and estimated masses are subject to systematic and random errors. A detailed study of the core-temperature mass issue, however, is beyond the scope of this research.

\subsection{Surface brightness profiles}
In contrast to many globules at lower Galactic latitudes, B207 is readily discernible through its surface brightness at optical wavelengths (Fig. \ref{fig:b207}). In the globule's optically thin outer portions, this surface brightness is proportional to the line-of-sight optical depth. Mapping the optical surface brightness of globules provides, therefore, an excellent means of tracing the optical depth distribution in the more transparent outer parts of a globule like B207. With additional information on the line-of-sight ISRF and the intensity of the diffuse galactic light (DGL) in the sky adjacent to the globule, it is also possible to estimate the globule's dust albedo at different wavelengths.

\begin{figure}
\includegraphics[width=0.50\textwidth]{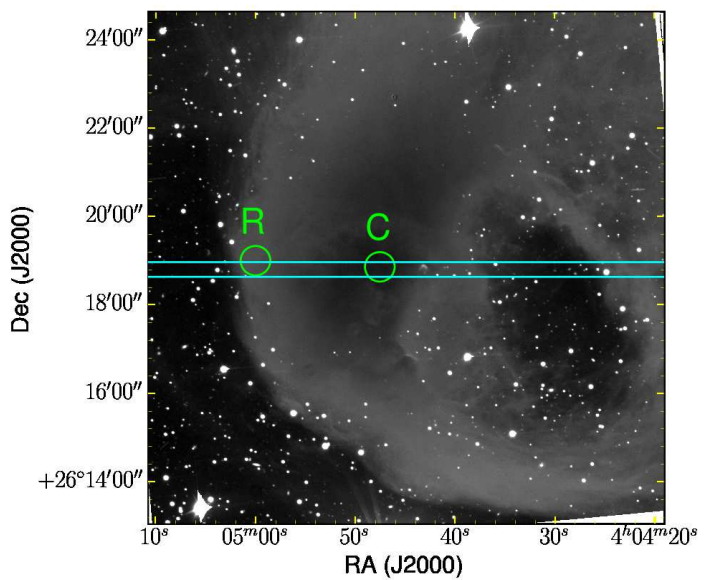}
\caption{Position of east-west horizontal cut region across B207 on an R-band image for surface brightness analysis. The core(C) and the rim(R) regions are indicated by the two circular apertures used to measure optical, NIR, MIR, FIR, and sub-mm intensities.}
\label{fig:bkgnd}
\end{figure}

We obtained surface brightness profiles of the globule B207 from the U, B, V, R, and I-band images. A horizontal east-west (E-W) cut, shown in Fig. \ref{fig:bkgnd}, was made to construct the surface brightness profiles. Individual surface brightness values were averaged over a box of size 5$\arcsec\times$20$\arcsec$, perpendicular to the E-W cut, oriented along the north-south (N-S) direction, yielding sky subtracted intensities in $erg/cm^{2}/s/A/sr$. To measure intensity profiles at 3.4 and 350 $\mu$m we used box sizes of 15$\arcsec\times$ 20$\arcsec$ and 25$\arcsec\times$ 25$\arcsec$, greater than their respective point spread function. Table \ref{table:calibration} lists the value of the conversion factor from $count/s/pixel$ to $erg/cm^{2}/s/A/sr$ in the U, B, V, R, and I bands, calibrated by measuring the flux from a Hubble standard star GD71.

\begin{table}
\caption{Photometric calibration for one count/s/pixel.}
\begin{tabular}{ccc}
\hline
Band & Effective wavelength ($\lambda$) & Intensity\\
name & (nm) & ($10^{-7} erg/s/cm^{2}/A/sr$)\\
\hline
U & 366 & 9.68\\
B & 428 & 1.63\\
V & 537 & 1.01\\
R & 633 & 0.683\\
I & 806 & 0.555\\
\hline
\end{tabular}
To calculate the surface brightness, we used the conversion of 1 pixel$^{2}$ = 1.354$\times$10$^{-12}$ steradian. The wavelength listed for each band is the arithmetic mean of the two wavelengths at which transmission is half of the maximum. The transmission curves for each band can be found at $www2.lowell.edu/rsch/LMI/specs.html$.
\label{table:calibration}
\end{table}

\begin{figure*}
\includegraphics[width=1.00\textwidth]{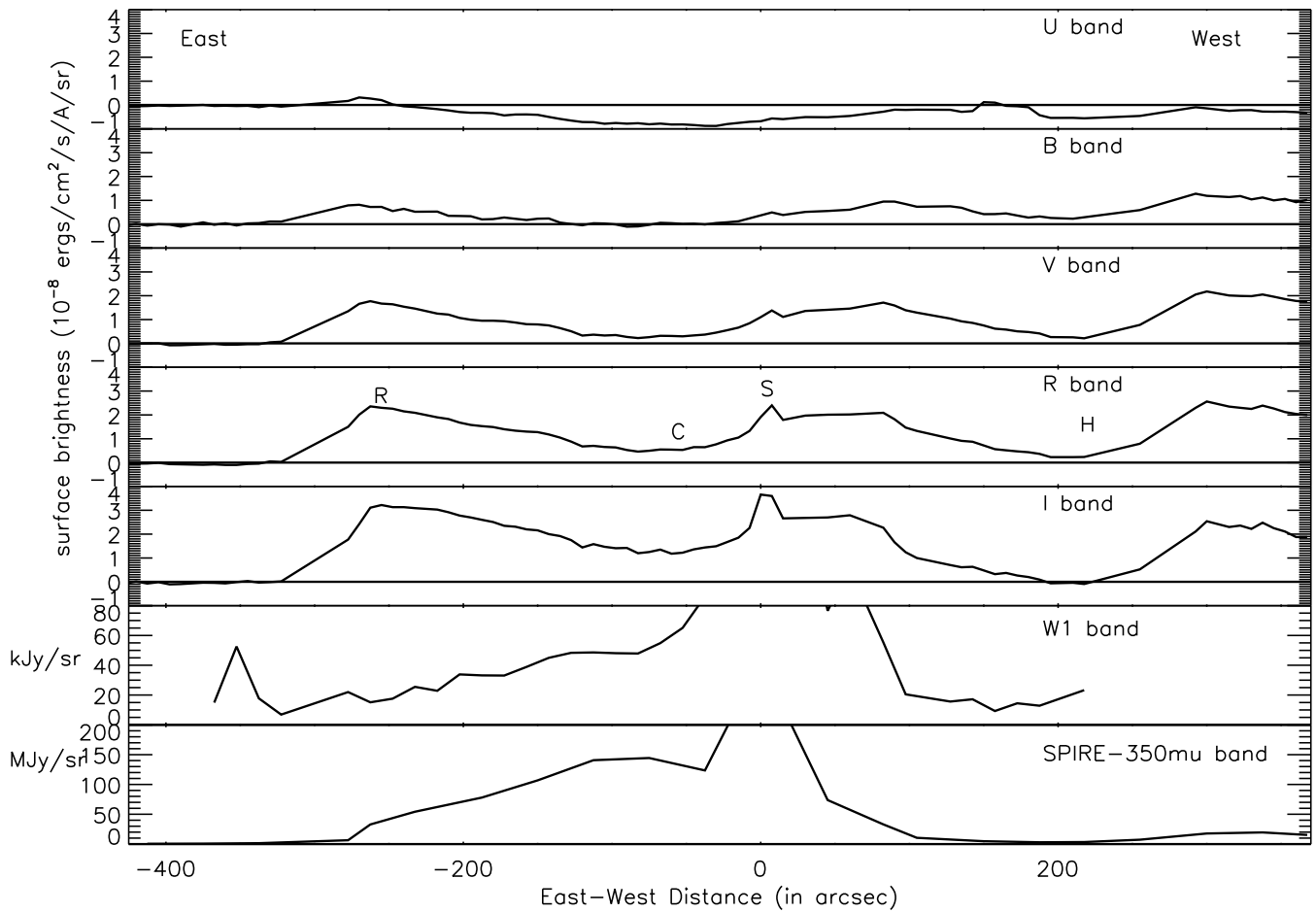}
\caption{Sky-subtracted surface brightness for B207 in U, B, V, R, I, WISE1-3.4 $\mu$m, and SPIRE-350 $\mu$m bands. The intensities in optical, 3.4 $\mu$m, and 350 $\mu$m are in ergs/cm$^{2}$/s/A/sr, kJy/sr, and MJy/sr, respectively. The east-west direction is marked in the U-band plot starting from east (left) to west (right). The position of the eastern bright rim, core, protostar IRAS 04016+2610, and the transparent hole region are marked as R, C, S, and H, respectively. The east-west distance is measured from the centre of the box close to the protostar, where the zero position corresponds to R.A. = $4^{h}4^{m}43.5^{s}$ , Dec = $+26^{\circ}18\arcmin50\arcsec$}
\label{fig:sb}
\end{figure*}

The surface brightness profiles for the five LMI bands along with WISE1 (3.4 $\mu$m) and 350 $\mu$m FIR SPIRE bands are shown in Fig. \ref{fig:sb}. At the eastern rim, the optical surface brightness rises sharply, reaching a peak almost exactly at the same spatial location for all the observed optical wavelengths. Given that the peak surface brightness at any given wavelength occurs at almost the same (wavelength-dependent) optical depth, this finding demonstrates that the line-of-sight optical depth increases sharply near the position of the brightness peaks. This suggests that the material is compressed from the eastern side of the globule, resulting in a steeper density gradient there than seen at other rim regions of B207. This steeper density gradient extends all the way to the center of B207, as is evident from the plot of the column density in Fig. 4.

At the core position, the surface brightness increases with increasing wavelength. The intensity of the core region is lower than the background sky in U and almost equal to the sky in B. The region west of the protostar, appearing as a hole in the globule, displays an intensity value greater than the clear background sky in B, V, and R bands, implying the presence of some diffuse dust. The flux of the protostar increases from undetectable at U to very bright at I and WISE1 bands, consistent with its SED peaking at $\sim 50$ $\mu$m \citep{Furlan08}.

The surface brightness distribution of B207 is characterized by a bright outer rim surrounding a dark core. This pattern is the consequence of a strongly forward-throwing phase function of the scattering dust, which controls the transfer of the illuminating ISRF through the object \citep{Witt74, Witt90}. High-latitude globules like B207 are detectable by their scattered-light surface brightness, because the DGL arising in the surrounding diffuse interstellar medium has a lower surface brightness due to its much lower line-of-sight optical depth, thus providing the contrast necessary for visibility. At low Galactic latitudes, the surface brightness of the diffuse galactic light (DGL) and of the bright outer rim of globules are more nearly the same, thus erasing this contrast. Such globules are then recognized simply as dark cores seen against a brighter background, the U-band surface brightness profile approaches this condition.

Given the extreme asymmetry of the scattering phase function, the globule's surface brightness is determined almost exclusively by the intensity of the ISRF seen on the far side of the globule from directions within a few degrees around the line-of-sight from the observer to the globule. The surface brightness at the side facing the observer is determined by the transfer of scattered radiation through different sections of the globule. The actual intensity is proportional to the product of two probabilities: the probability of scattering, measured by $a(1-e^{-\tau})$, and the probability that once-scattered radiation escapes through the front surface of the globule, measured by $e^{-\tau_{abs}}$. Here, $a$ is the dust albedo, $\tau$ is the line-of-sight extinction optical depth through the globule, and $\tau_{abs}$ is the typical absorption optical depth for the average once-scattered photon.  A suitable approximation is $\tau_{abs} \approx 0.5(1-a)\tau$ \citep{Witt08, Witt85}. The bright outer rim of an externally illuminated globule occurs where the product of these two probabilities reaches a maximum. For the high dust albedos typical for globules \citep{Witt90}, this maximum occurs at line-of-sight optical depths near $\tau = 1.6$, occurring at about $\sim$-270$\arcsec$ from the protostar in B207, as seen in Fig. \ref{fig:sb}.

\subsection{Dust characteristics}
\subsubsection{Scattered light intensities}
The maximum surface brightness of B207 observed in the different LMI bands may be used to constrain the albedo of the dust present in the outer portion of this globule.  A convenient tool is the relation between the ratio of maximum surface brightness to the intensity of the illuminating radiation field and the dust albedo, as found from Monte Carlo radiative transfer simulations by \citet{Witt74}. In order to employ this tool, we need to correct the observed maximum surface brightnesses in the five LMI bands for the background DGL intensity still present in the intensity of the sky adjacent to B207, and we also need the value of the ISRF in the direction of B207. It is this particular intensity that is crucial for determining the surface brightness of B207 because of the strongly forward-throwing phase function of the dust.

The sky intensity adjacent to B207 consists of foreground components stemming mainly from terrestrial airglow, scattering in the Earth's atmosphere, and zodiacal light, I$_{fg}$, and a background component consisting of DGL, I$_{DGL,sky}$, arising from scattering of the ISRF by diffuse Galactic dust along the line of sight. Hence,
\begin{equation}
I_{sky} = I_{fg}+I_{DGL,sky}.
\end{equation}
The globule intensity is composed of ISRF photons scattered by dust in B207, I$_{sca}$, plus the same foreground intensity, I$_{fg}$, seen in the adjacent sky and a fraction of the DGL component that is being transmitted in optically thin outer portions of the globule. The ISRF intensity contains contributions from both stars and DGL arising at distances beyond that of B207:
\begin{equation}
I_{globule} = I_{fg}+I_{sca}+I_{DGL,sky} e^{-\tau'},
\end{equation}
where $\tau'$ is the line-of-sight extinction optical depth through the globule at the position, where the intensity is being measured. This assumes that no significant DGL contribution arises in the space between us and B207. This assumption is supported by two facts. The core of B207 in U has a substantially lower surface brightness than that of the adjacent sky, consistent with the assumption that all DGL arises at distances beyond that of B207. Furthermore, recent work by \citet{Green15} places most of the interstellar dust in the direction of B207 at a distance of 250 pc and beyond, while the distance to B207 is about 140 pc. Thus, subtracting the sky intensity from the globule intensity leaves us with
\begin{equation}
I_{globule}-I_{sky} = I_{sca}+I_{DGL,sky} (e^{-\tau'}-1).
\end{equation}
At the position of the maximum surface brightness (SB$_{max}$) the line-of-sight optical depth ($\tau'$) is $\sim$1.5 (Fig. 6 of \citealt{Witt82}) and there we find
\begin{equation}
I_{globule}-I_{sky}+I_{DGL,sky} \times 0.78 = I_{sca}.
\label{eqn:eqndgl}
\end{equation}

To find the absolute surface brightness of B207, I$_{sca}$, we need to add the intensity of the adjacent sky DGL fraction to the intensities displayed in Fig. \ref{fig:sb}, which are background sky-subtracted surface brightness values. The foreground components, contributing equally to both the sky and globule intensities, have been eliminated by the original sky subtraction and are of no further concern.

We estimate the DGL intensities in the sky adjacent to B207 by first finding the optical depth of the dust in the sky region. The all-sky map of \citet{Schlegel98} yields a value of E(B - V) $\sim$ 0.197, and with the re-normalization by \citet{Schlafly10} and \citet{Schlafly11} yields a more probable value of E(B-V) $\sim$ 0.169. With a value of R$_{V}$ = 3.1 for average diffuse Galactic dust, we find A$_{V}$ = 0.53 mag. Scaling this value with the extinction cross sections per H nucleon (C$_{ext}$/H) as a function of wavelength from \citet{Draine03}\footnote{\url{www.astro.princeton.edu/~draine/dust/dustmix.html}} for ISM dust (R$_{V}$ = 3.1), we find the optical depth values listed in Column 3 of Table \ref{table:dust}. 

Noting that these optical depths are within the optically thin regime, we use the radiative transfer approach based on single scattering \citep[Equation 4;][]{Witt08} to estimate the corresponding DGL intensities. For this, we use the ISRF intensities of \citet{Porter05} at the globule's position, listed in Column 2 of Table \ref{table:dust}, and the dust albedos from \citet{Draine03} for R$_{V}$ = 3.1 Milky Way dust, listed in Column 4 of Table \ref{table:dust}. The resulting DGL intensities are listed in Column 6 of Table \ref{table:dust}.
\begin{table*}
\centering
\begin{minipage}{180mm}
\caption{Dust characteristics}
\begin{tabular}{ccccccccc}
\hline
Filter & ISRF & $\tau$ & a & $\tau_{abs}$ & $I_{DGL,sky}$ & $SB_{max}$ & $\frac{(0.78 \times I_{DGL,sky}+SB_{max})}{ISRF}$ & a$_{c}$\\
\hline
U & 4.77 & 0.721 & 0.625 & 0.135 & 1.338 & 0.259 & 0.273 & 0.54$\pm$0.05\\ 
B & 5.55 & 0.624 & 0.648 & 0.110 & 1.497 & 0.839 & 0.361 & 0.65$\pm$0.05\\
V & 6.41 & 0.485 & 0.671 & 0.080 & 1.528 & 1.811 & 0.468 & 0.74$\pm$0.05\\
R & 6.80 & 0.402 & 0.676 & 0.065 & 1.425 & 2.361 & 0.511 & 0.77$\pm$0.03\\
I & 6.37 & 0.259 & 0.656 & 0.045 & 0.911 & 3.272 & 0.625 & 0.84$\pm$0.03\\
\hline
\end{tabular}
\\
a. The values of ISRF, DGL, and SB$_{max}$ are in units of 10$^{-8} erg/s/cm^{2}/A/sr$. The ISRF for different wavelengths are from \citet{Porter05}.\\
b. $\tau$ and $a$ are the extinction optical depth and albedo of the dust in the sky adjacent to B207 assuming R$_{V}$ = 3.1. The albedo values are from \citet{Draine03}.\\
c. $\tau_{abs}$ is the average absorption optical depth of photons after their first scattering, $\tau_{abs} = 0.5(1-a)\tau$\\
d. $I_{DGL,sky}$ is the diffuse galactic light adjacent to B207, derived by using $a I_{ISRF} (1-e^{-\tau}) e^{-\tau_{abs}}$\\
e. $SB_{max}$ is the peak intensity of the globule after subtracting the nearby background sky intensity.\\
f. The quantity (0.78$\times$DGL+SB$_{max}$)/ISRF is the ratio of the maximum intensity at the rim of the globule to the intensity of the illuminating interstellar radiation field.\\
g. The albedo of the cloud, a$_{c}$, evaluated from the corresponding ratio (0.78$\times$I$_{DGL}$+SB$_{max}$)/ISRF assuming the forward scattering asymmetry parameter, $0.6 \leq g \leq 0.9$ from Fig. 6, \citet{Witt74}.
\label{table:dust}
\end{minipage}
\end{table*}

In order to arrive at the background-corrected absolute surface brightness values for the maximum surface brightness, we add this DGL component to the sky-subtracted maximum surface brightness listed in Column 7 of Table \ref{table:dust}. The ratio of the corrected absolute maximum surface brightness to the ISRF listed in Column 8 is the quantity to be compared directly with the model predictions in Fig. 6 of \citet{Witt74}. The resulting estimates for the dust albedo, $a_{c}$, in B207 are shown in the last Column of Table \ref{table:dust} with their estimated uncertainties. The albedo values were calculated under the assumption that the phase function asymmetry, g, is close to 0.9, consistent with the results found in other dark nebulae \citep{Gordon04}. \citet{Henyey41} introduced the phase function asymmetry parameter, g, in the range $-1\leq g \leq +1$, from back scattering to isotropic scattering (g = 0) to forward scattering. A change in the assumption of phase function asymmetry to g $\approx$ 0.7 causes a small reduction, about 0.02, in the estimated albedo values \citep{Gordon04}.

Two facts stand out from this analysis. First, the globule albedo values are generally larger than those of the dust in the adjacent diffuse dust background. Second, the albedo increases with wavelength, reaching its highest value of about 0.84 in the I band. These characteristics are typical for size distributions of grains that include micron-size grains, which are significantly larger than the largest grains in diffuse ISM dust, as shown for example by \citet{Kim96}. The dust albedo values are comparable with those found in other studies of globules and dark nebulae. \citet{Fitzgerald76} evaluated the dust albedo in B band, $a = 0.70\pm0.08$, for the Thumbprint nebula. The Coalsack and Libra clouds have a dust albedo of $\sim0.6$ for g $\sim0.8$ in U, B, and V bands \citep{Mattila70}. However \citet{Witt90} found that the albedo value decreases from 0.80 at 469 nm to 0.58 at 856 nm for the Bok globule CB4, but this globule has a much smaller central optical depth compared to B207.   

\subsubsection{Coreshine effect in NIR bands}
\begin{figure}
\centering
\includegraphics[width=0.45\textwidth]{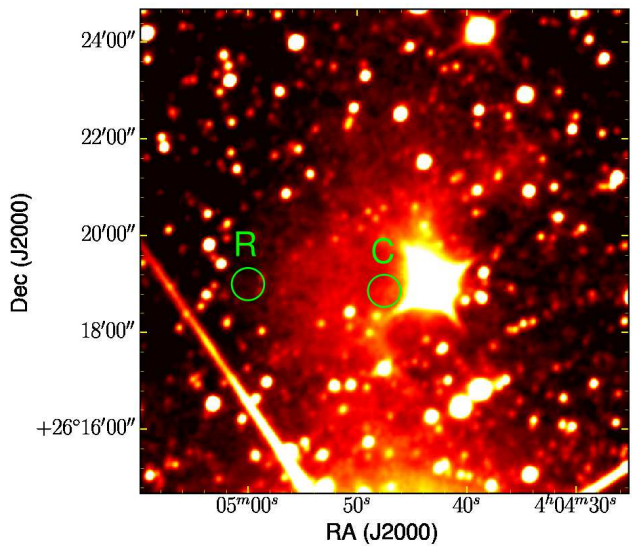}
\caption{$10\arcmin\times10\arcmin$ WISE W1 image of B207 at 3.4 $\mu$m. The protostar IRAS 04016+2610 is at the centre of the image. B207 is exhibiting the coreshine effect by showing a clear excess of 3.4 $\mu$m emission. The protostar IRAS 04016+2610 is at the centre of the image. B207 is exhibiting the coreshine effect by showing a clear excess of 3.4 $\mu$m. The coreshine effect is due to the scattering by large grains inside the globule. The two circular regions marked on the image are the core (C) and rim (R) regions, where we measured optical-, near-, mid-, and far-infrared intensities.}
\label{fig:b207wise1}
\end{figure}

The 3.4 $\mu$m WISE W1 image of B207, shown in Fig. \ref{fig:b207wise1}, reveals the clear presence of coreshine. The coreshine phenomenon can be explained by near-infrared (NIR) radiation from the ISRF being scattered by micron-size grains. The dense cores of the globules are cold and thus provide favourable physical environments for the grains to acquire ice mantles, coagulate, and grow in size to exhibit the coreshine effect \citep{Hirashita13, Andersen14}. The intensity of the coreshine depends on the incident radiation, the extinction of the background radiation, the grain properties, and the core properties \citep{Steinacker14b, Lefevre14}. In the case of B207, the presence of the nearby protostar IRAS 04016+2610 at a projected offset distance of 8400 au may result in an enhancement of the mid-IR radiation field that illuminates the core. This possibility will be discussed in greater detail in Sect. 4.5.

We measured the intensity of the coreshine in the core and rim regions of the globule B207, marked by two circles in Fig. \ref{fig:b207wise1}. We list the background subtracted intensities for these two regions in Table \ref{table:intensity} at different wavelengths from optical-, near-, mid-, and far-infrared wavelengths using DCT, 2MASS, WISE, and Herschel images, respectively. In the mid- and near-infrared wavelengths the conversion from DN to Jy units is adopted from the WISE Preliminary Data Release\footnote{\url{wise2.ipac.caltech.edu/docs/release/prelim/expsup/sec2_3f.html}} and using the flux for zero-magnitude zero point conversion values for 2MASS\footnote{\url{www.ipac.caltech.edu/2mass/releases/allsky/doc/sec6_4a.html}}, respectively. We estimated the ISRF at NIR wavelengths 1.25--4.62 $\mu$m by scaling the \cite{Mathis83} ISRF at 10.0 kpc to match the optical ISRF from \cite{Porter05} at optical wavelengths, and we used these values to estimate the DGL at NIR wavelength by the method described in Sect. 3.4.1.

We found B207's core intensity excess above sky at 3.4 $\mu$m to be 0.066 MJy/sr. This surface brightness value of the B207 core is comparable to other star forming cores found in the Taurus-Perseus region \citep{Lefevre14}. Theoretical modelling by \citet{Steinacker14b} predicted a somewhat lower value of $\sim0.06$ MJy/sr for the surface brightness at a slightly longer wavelength of 3.6 $\mu$m. The core intensity at 3.36 $\mu$m listed in Table \ref{table:intensity} is slightly higher than the corresponding W1 surface brightness shown in Fig. \ref{fig:sb}. The intensities in Tab. \ref{table:intensity} were measured using a 20$\arcsec$ radius circular aperture while the intensities in Fig. \ref{fig:sb} were obtained with a four-times smaller size rectangular box aperture. As a result, the intensity at 3.36 $\mu$m in Table \ref{table:intensity} was more strongly contaminated by the presence of the protostar. The relatively high observed surface brightness at the core of B207 suggests that the nearby class I protostar, IRAS 04016+2610, is in fact contributing to the core illumination at 3.4 $\mu$m to a small degree.

\begin{table}
\caption{DGL and optical-IR-submm sky-subtracted intensities for B207}
\begin{tabular}{ccccc}
\hline
Wavelength & Core & Rim & DGL & f$_{DGL}$\\
$\lambda$ ($\mu$m)& MJy/sr & MJy/sr & MJy/sr & MJy/sr\\
\hline
0.43 &-5.43E-5 & 4.02E-3 & 9.14E-3 & 3.44E-3 \\
0.54 & 2.44E-3 & 1.55E-2 & 1.47E-2 & 4.17E-3 \\
0.63 & 7.89E-3 & 3.05E-2 & 1.90E-2 & 4.11E-3 \\
0.81 & 2.74E-2 & 6.82E-2 & 1.97E-2 & 2.00E-3 \\
1.25 & 9.06E-2 & 1.14E-1 & 1.27E-2 & 9.76E-4 \\
1.65 & 9.90E-2 & 1.62E-1 & 7.22E-3 & 3.38E-4 \\
2.17 & 6.66E-2 & 1.10E-1 & 3.71E-3 & 1.37E-4 \\
3.36 & 6.64E-2 & 2.61E-2 & 7.89E-4 & 1.79E-5 \\
4.62 & 1.02E-1 & 2.24E-2 & 1.77E-4 & 3.16E-6 \\
100  & 9.02E+0 & 9.53E+0 & --- & --- \\
160  & 6.94E+1 & 3.75E+1 & --- & --- \\
250  & 1.55E+2 & 5.80E+1 & --- & --- \\
350  & 1.45E+2 & 4.14E+1 & --- & --- \\
500  & 8.20E+1 & 1.99E+1 & --- & --- \\
\hline
\end{tabular}
\\
a. The uncertainties in the observed intensities and the DGL are about 10$\%$ and 25$\%$, respectively. However, for H (1.65$\mu$m) and K$_{s}$ (2.17$\mu$m) bands the uncertainties are higher than the observed sky subtracted intensities.\\
b. The centre of the core region is R.A. = $4^{h}4^{m}47.539^{s}$ , Dec = $+26^{\circ}18\arcmin51.66\arcsec$. \\
c.The centre of the rim region is R.A. = $4^{h}5^{m}00.00^{s}$ , Dec = $+26^{\circ}19\arcmin00.00\arcsec$. \\
d. The intensities are measured using the optical, NIR, MIR, and FIR wavelengths from DCT, 2MASS, WISE, and Herschel PACS and SPIRE images.\\
\label{table:intensity}
\end{table} 

\subsubsection{THEMIS model}
The variations in the dust SED due to temperature, far-IR/submm opacity and spectral index, in both the diffuse and dense ISM were reproduced by the THEMIS model \citep{Kohler15}. Within the context of the THEMIS model, grains up to a radius of 20 nm consist purely of aromatic-rich H-poor amorphous carbon, a-C, whereas bigger grains have a core/mantle (CM) structure, where the core consists of amorphous silicate (Mg-rich forsterite and enstatite-normative compositions) or of aliphatic H-rich amorphous carbon. \citet{Kohler15} assumed that the dust properties change with increasing local density through accretion and coagulation. The model consists of several dust population mixtures, labelled according to \citet{Kohler15}:\\
CM (core mantle): no evolution of diffuse ISM dust;\\
CMM (core mantle mantle): dust grains with a second H-rich carbon mantle, present in the molecular cloud edges;\\
CMM+AMM (core mantle mantle + aggregates): CMM grains coagulate to form aggregates in the outer molecular core regions;\\
CMM+AMMI (core mantle mantle + aggregates with ice): similar to the previous case but includes the formation of ice mantles on the aggregates in the dense core regions.

The model in \citet{Ysard16}, applicable to the globule B207,  assumes a dense spherical cloud with central number density n$_{c}$ = 10$^5$ cm$^{-3}$ and column density N(H) = 2.9 $\times$ 10$^{22}$ cm$^{-2}$ with a gas mass of 8 M$_{\sun}$. The cloud is illuminated by the interstellar radiation field and consists of CM, CMM, CMM+AMM, and CMM+AMMI dust populations. The number density at the core of B207 is 5.6 times higher than assumed by the \cite{Ysard16} model and would increase even further, if the actual core temperature is lower than that derived by the Abel Inversion technique. However, as is shown in Fig. 8d of \cite{Ysard16}, a change in central number density from 10$^{5}$ to 10$^{6}$ H/cm$^{3}$ decreases the model intensity of CMM+AMMI grains by a factor less than two at optical, and by a still smaller factor at NIR wavelengths. The model cloud of \citet{Ysard16} is sufficiently similar to B207 in terms of density, column density, and mass,to allow meaningful comparisons between the observed B207 scattering and emission intensities with the predicted model intensities.

\begin{figure}
\includegraphics[width=0.48\textwidth]{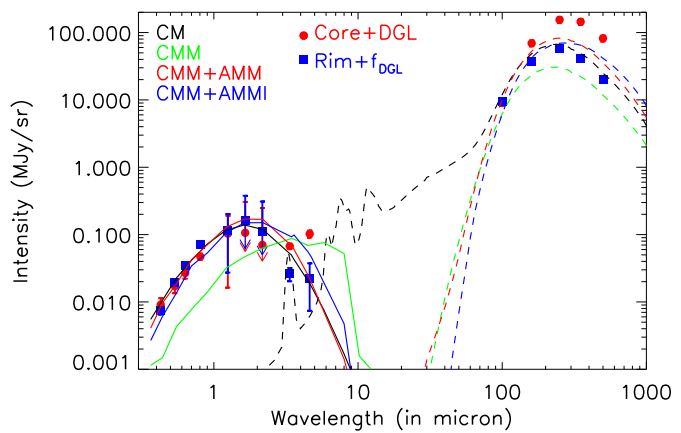}
\caption{Core+DGL (red circles) and rim+f$_{DGL}$ (blue squares) intensities of B207 as a function of wavelength. The solid and dashed lines in the near-to-MIR and FIR are the scattering and emission spectra of dust, respectively. Different line colours (black, green, red, and blue) represent CM, CMM, CMM+AMM, and CMM+AMMI grains, respectively.}
\label{fig:ysard}
\end{figure}

For this purpose, we selected two regions at the core and the rim, marked by green circles in Fig. \ref{fig:b207wise1}, to measure and compare our intensities with the THEMIS model. In Table \ref{table:intensity} we list our sky-subtracted intensities for these two regions at different wavelengths. The projected distances of the protostar in the sky plane from the centre of our marked core and rim regions are 8400 and 32000 au, respectively. The THEMIS model assumes no dust in the sky background, and hence no sky DGL is included in the model. Our sky-background subtraction  results in a lowering of the measured intensities at the core and rim regions by amounts equal to the intensity of DGL; hence, we need to add the sky DGL to our measured intensities before comparing them with the THEMIS predicted intensities. In contrast to the optically thick core, the rim is optically thin. This allows the scattered fraction of DGL to be transmitted at the rim region and hence, only the absorbed fraction of total DGL, f$_{DGL}$, needs to be added. This fraction is calculated using $(1-e^{-\tau_{abs}'}) \times$ DGL, where $\tau_{abs}'$ is the absorption optical depth of the cloud estimated from the relation $\tau_{abs}' = (1-a_{c}) \times \tau'$. We estimated $\tau'$ = 1.3 in the V band and scaled this value to applicable optical depths for the other bands by using the dust model of R$_{V}$ = 5.5. The values of DGL and f$_{DGL}$ for the direction of B207 are listed in Table 5 for wavelengths from the optical to MIR. In Fig. \ref{fig:ysard} we plot our intensities after having added the DGL and the f$_{DGL}$ for the core and rim, respectively, along with the THEMIS model values of CM, CMM, CMM+AMM, and CMM+AMMI from \citet{Ysard16}.

\section{Results and discussions}
\subsection{Physical properties of the globule}
\subsubsection{Temperature}
Assuming a single temperature black body to fit the SED at FIR wavelengths, the dust temperature at the core position is estimated to be about 11 K. The core temperature is about 3 K less than the dust temperature in the outer warm regions of the globule. With the Abel Inversion technique the measured dust temperature at the core is found to be about 1.7 K less than the SED-measured temperatures. This difference is in agreement with the results obtained by \citet{Roy14} for the globule B68 and L1689. The SED technique uses the measured FIR intensities averaged over the line of sight. The outer area of the projected core region is at higher temperatures than the inner region, which is partially shielded against the ISRF and other sources responsible for heating the dust and gas. At the outer regions, towards the rim of the globule, the measured temperatures using both methods, SED fitting and Abel inversion technique, yield similar values for the dust temperatures. However, we would like to caution our readers that recent studies have shown temperatures in the core can be as low as in the range 6--8 K.

\subsubsection{Density and mass}
SED analysis:\\
The column density, N(H), estimated at the central dense core with the SED technique, is 7.2 $\times$ 10$^{22}$ cm$^{-2}$ (Table \ref{table:firint}). Assuming the core radius of 9800 au, as discussed in Sect. 3.2, this column density corresponds to an average number density n(H) = 2.5 $\times$ 10$^{5}$ cm$^{-3}$. Using the relation between the column density and A$_{V}$ given by \citet{Bohlin78} with an assumed value of R$_{V}$ = 5.5 for molecular clouds, the optical extinction, A$_{V}$, in the core is 48 mag. The column density is reduced by about 30 times near the rim region, yielding E(B-V)$\sim$0.3 towards the rim of B207. The column density gradient south-east of the core along our cut is steeper when compared to that in the north-west (Fig. \ref{fig:tempden}), where we observe diffuse gas, shaped like a tail morphology, stretching over a pc in length distance toward the north-west. The gas appears compressed from the east and south-east, which may be due to external pressure; however, no likely source of any shocks is observed in the vicinity of B207.\\
\\
Abel inversion analysis:\\
Using the Abel inversion technique, the number density measured at the central core region is about 5.6$\times$10$^{5}$ cm$^{-3}$, a factor of 2.2 higher than estimated from the SED technique. This value is slightly higher than the ratio of estimated column densities from Abel inversion to the SED method (Table \ref{table:firint}), which may be due to the temperature averaging to estimate T$_{abel}$ for each box region. The estimated exponent of the number density distribution from the Abel Inversion technique in the inner portion of the core is about -0.85, n$_{H}$ $\propto$ r$^{-0.85}$, consistent with the 450 and 850 $\mu$m SMA results of \citet{Hoger00}. We measured the gas mass in each rectangular box region and summed them. The resulting mass of the core was estimated to be 5.6 M$_{\odot}$ or 3.2 M$_{\odot}$, when assuming either $\kappa_{500}$ for AMM/AMMI dust grains \citep{Kohler15} or for grains with thin ice mantles \citep{Ossenkopf94}, respectively. Our measured intensities follow the AMM/AMMI evolved dust grain type model; hence, the core mass estimate of 5.6 M$_{\odot}$ seems more plausible. With the measured 500 $\mu$m FIR integrated intensities over the 12.5$\arcmin\times$12.5$\arcmin$ FOV, shown in Fig. \ref{fig:b207}, the estimated gas mass is about 17 M$_{\odot}$, assuming an average temperature of 12 K and $\kappa_{500}=2.85$ cm$^{2}$/g for AMM and AMMI type dust grains \citep{Kohler15}. The measured temperature, column density, number density, and mass of B207 are very similar to values found in typical Bok globules. However, recent studies have indicated the presence of very cold dust ($<$ 8 K) in molecular cores, potentially missed by the Abel inversion method. Ongoing efforts are pursued to estimate the cold dust temperature and mass fraction present in the central core regions of Bok globules.

\subsection{Surface brightness in core and rim}
The surface brightness profiles in U, B, V, R, and I bands display a bright rim surrounding a dark core, with increasing core surface brightness at longer wavelengths. The darkness of the core is due to the inefficiency of back scattering with a highly forward directed scattering phase function. The increase in dust albedo, in addition to an increase in the ISRF intensity at longer wavelengths, causes the core intensity to increase in redder bands. Given the forward directed phase function and the high photon escape probability in the optically thin rim, the surface brightness is greatly enhanced. Hence, the rim is brighter than the core.  

The adjacent sky is darker mainly because of lower optical depths, declining with increasing wavelengths. In addition, the albedo of the dust grains in the sky region near to B207 are lower compared to the dust in the globule, and this difference increases with increasing wavelengths (Table \ref{table:dust}). As a result, the contrast between the rim and the nearby sky region increases with wavelength, causing the rim intensity to increase significantly above the sky background, while at the shortest wavelength in the U band the intensities of the rim and the sky become almost equal, making the rim nearly undetectable. 

\subsection{Grain growth and age estimate}
The dust albedo $a_{c}$ of the cloud listed in Table \ref{table:dust} increases with wavelengths from about 0.54 in the U band to around 0.84 in the I band. The increase of the albedo with increasing wavelength suggests an increase in the grain size inside the globule compared to the dust in the diffuse ISM \citep{Kim96, Draine03a}, where the dust albedo is $\sim$0.65, as listed in Column 4 of Table \ref{table:dust} for R$_{V}$ = 3.1. This suggests that the cloud possesses conditions favourable for grain growth. The observed phenomenon of coreshine in the WISE W1 band also indicates the presence of large grains in the core of B207. With a core density of 10$^{5}$ cm$^{-3}$, the time required for grains to coagulate and grow to micron size is about 10$^{5}$--10$^{6}$ years \citep{Hirashita13, Ysard13, Steinacker14a}. The presence of the Class I protostar, IRAS 04016+2610 \citep{Yen14}, also indicates an age of the globule in the range of a few$\times10^{5}$ years.

\subsection{Type of dust grains in the globule B207}
\begin{figure}
\includegraphics[width=0.5\textwidth]{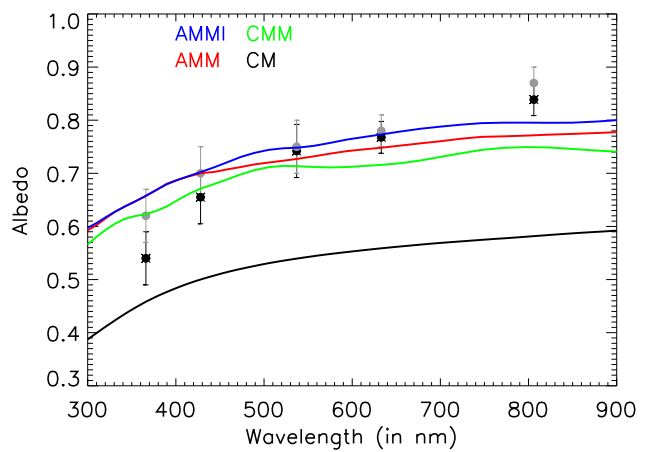}
\caption{Our derived rim dust albedos (black circles) are calculated using the ratio (0.78$\times$I$_{DGL}$ + SB$_{max}$)/ISRF and plotted along with the THEMIS model values for the full size distribution of CM, CMM, AMM, and AMMI grains. The grey circles are dust albedos estimated assuming no reduction in DGL (using the ratio (I$_{DGL}$ + SB$_{max}$)/ISRF). A reduction of 13$\%$ and 7$\%$ in albedo values are seen in U and B bands, respectively, with insignificant changes at higher wavelengths for a 22$\%$ reduction in DGL intensity.}
\label{fig:jones}
\end{figure}

We compared the albedo of the dust grains measured at the rim of the globule with the THEMIS albedos (Fig 11 of \citealt{Jones16}). Our measured albedos are in agreement with the THEMIS model for evolved dust grains: CMM, CMM+AMM, and CMM+AMMI, as shown in Fig. \ref{fig:jones}. The possibility of normal CM type dust grains can be ruled out. By assuming that the entire DGL in the sky adjacent to B207 arises at distances beyond B207, our analysis approach maximizes the amount of DGL to be added to the observed U, B, V, R, I surface brightness and, hence, results in upper limits to the actual albedo values. The 3-D dust map of \citet{Green15} shows less than 20$\%$ of the line-of-sight dust distribution in the direction of B207 at distances less than 140 pc, the estimated distance of B207. By accepting that 22$\%$ of the background DGL is being transmitted at the position of maximum surface brightness in the rim (Eq. \ref{eqn:eqndgl}), the DGL that must be added is reduced by an equivalent amount. The consequence of this step is a reduction in the derived albedo values from the absolute upper limits by small amounts, as illustrated in Fig. \ref{fig:jones}. This would still leave them in fair agreement with the model albedos shown in Fig. \ref{fig:jones}, and would bring down the albedo for the I-band into more acceptable agreement. We do note, however, that the gradient in the albedo increase with wavelengths as derived from our observations of B207 and is noticeably steeper than that of any of the model predictions. This could indicate a difference in the optical constants adopted for the model calculations from those in the actual grains as well as differences between the true size distribution and those adopted in models.

In Fig. \ref{fig:ysard} we compare the predicted SED of the cloud core model of \citet{Ysard16} with our observations. There is excellent agreement in the optical and near-IR range, where intensities are the result of scattering in the outer portions of the cloud and, thus, are insensitive to the column density of the optically thick core. In the far-IR and sub-mm region, our observed core intensities are higher by about a factor of two and the rim intensities are lower by about the same factor, when compared to the THEMIS model with evolved, larger dust grains. This is not unexpected for this wavelength range, where the core is optically thin. Our derived column density at the core of B207, N(H) = 7.2$\times$ 10$^{22}$ cm$^{-2}$, is about 2.5 times larger than the core column density modelled by \citet{Ysard16}. The observed core intensity should therefore be higher by a similar amount. Similarly, the column density at our rim position is smaller than that adopted by \citet{Ysard16}, with a corresponding difference in the far-IR and sub-mm intensities. Given that the THEMIS cloud model relies on heating by the external ISRF only, this agreement of our observations with the model also suggests that the presence of the class I protostar, IRAS 04016+2610, at a projected offset distance of 8,400 au from the core does not result in a measurable increase in the dust heating in the cloud core. This supports the finding by \citet{Launhardt13} that the presence of a protostar affects the physical conditions of the core if the star is within a distance of less than 5000 au. However, the protostar in B207 is likely to affect the observations in the mid-IR, as we will discuss in the following sub-section.

Given that the observational data for the optical and near-IR range shown in Figs. \ref{fig:ysard} and \ref{fig:jones} reflect conditions in the outermost layers of B207,  we can conclude that the dust grains in the bright rim of B207 are of CMM+AMM or CMM+AMMI type. \citet{Ford11} modelled B207's core with a canonical abundance of CO and found a significant depletion compared to the ISM value. This indicates freezing of CO and other major molecules like H$_{2}$O onto grains, supporting the possibility of AMMI type dust grains. In an externally illuminated, optically thick system such as B207, the estimated dust scattering characteristics at optical wavelengths apply only to the outermost optically thin layers because only those are being probed by photons incident from the external ISRF, limiting our estimation of dust albedo to the rim. In the core region of B207 we have no information on optical properties; we can only conclude that the dust grains are of CM, CMM+AMM, or CMM+AMMI type also from the agreement of the FIR and sub-mm data with the prediction from the THEMIS cloud model. In Sect. 4.6 we will discuss how coagulated grains from the inner core may be transported to the cloud surface.

\subsection{Effects of the protostar on core properties}
\begin{figure}
\includegraphics[width=0.5\textwidth]{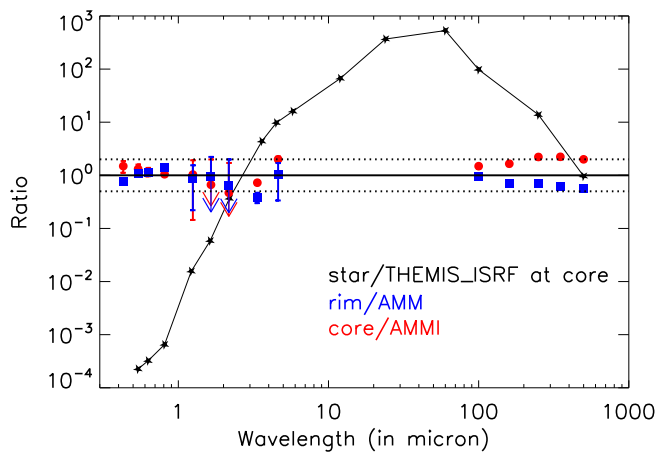}
\caption{Ratio of star intensity to the THEMIS ISRF at the core position, and ratio of the observed core and rim region intensity to the THEMIS AMMI and AMM models, respectively, as a function of wavelength. The dotted lines denote the limits for a factor of two uncertainty. The class I protostar intensity at the core is negligible at optical and NIR wavelengths but dominates at the mid-IR and far-IR wavelengths. The ratio of observed core and rim intensities to THEMIS model intensities for AMMI and AMM dust grains, respectively, shows agreement within a factor of two. The high column density at the B207 core results in FIR core intensities higher than the THEMIS AMMI predicted values by a corresponding factor.}
\label{fig:starisrf}
\end{figure}

The SED of the class I protostar, IRAS 04016+2610, has been modelled from the optical to the sub-mm wavelengths \citep{Gramajo10, Robitaille07} . We calculated the radiation field intensities due to this star at a linear offset distance of 8,400 au, that is, the projected core position, thus arriving at upper limits for the stellar flux that could affect the conditions at the core. The class I protostar may be positioned behind the core \citep{Brinch07} leading to an increase in the actual protostar distance. We compared the radiation densities due to the star to the radiation densities of the ISRF used in the THEMIS cloud model, both of them uncorrected for in-cloud extinction, with their ratio plotted as a continuous thin black line in Fig. \ref{fig:starisrf}. We also show for comparison in Fig. \ref{fig:starisrf} the ratio of observation to model data from Fig. \ref{fig:ysard}.

Figure \ref{fig:starisrf} clearly shows that the stellar contribution to the cloud illumination at optical and near-IR wavelengths is quite insignificant. This implies that the high albedo values found in the even more distant rim region are in no way affected by the presence of the star. Also, the star does not contribute significantly to the heating of the core at these shorter wavelengths. The IRSF photons in the optical and near-IR contain most of the energy of the ISRF and reach the core via diffusion through multiple scattering in a high-albedo dust environment. In the mid-IR, at wavelengths 3.6 $\mu$m and 4.5 $\mu$m, the radiation fields due to the star and due to the ISRF are of comparable magnitude. It is very likely, therefore, that illumination by the protostar is contributing to the observed coreshine at these wavelengths, as we suggested in Sect. 3.4.2. This is further supported by the extremely red colour of the observed coreshine over the range of these two wavelengths. The coreshine ratio, 4.5 $\mu$m / 3.6 $\mu$m, for the CM, AMM, and AMMI dust grains predicted by the THEMIS model are in the range 0.42--0.56, consistent with the results of \citet{Lefevre14} for starless cores in the Taurus-Perseus region of the sky. The observed coreshine ratio (4.6 $\mu$m / 3.4 $\mu$m) for the core of B207 is 1.54 (Table \ref{table:intensity}). This suggests that the coreshine at 4.6 $\mu$m is substantially dominated by the illumination from the protostar. Nevertheless, the presence of the coreshine at these wavelengths, regardless of the illumination source, is evidence of high dust albedos in the mid-IR, characteristic of large, evolved grains.

It is possible that illumination in the mid-IR by the protostar could contribute to the heating of dust grain in B207, in particular through absorption in the water band near 3.1 $\mu$m and the silicate band near 9.7 $\mu$m. This is not likely to be a significant contribution, because we were able to account for the higher observed far-IR and sub-mm intensities at the core position by noting that B207 has a core column density more than twice that of the THEMIS model cloud without requiring grains of higher temperatures than that expected from heating by the ISRF alone.

In the far-IR, the radiation field at the core of B207 originating with the protostar totally dominates over the ISRF. However, given that the dust albedo at such long wavelengths is essentially zero, there is no contribution to the far-IR core intensity from scattering. The contribution to the heating of grains at the core due to stellar far-IR flux is also insignificant, because the absorption cross sections of grains at these wavelengths are lower than the cross sections at UV, optical, and near-IR wavelengths by four to five orders of magnitude, which more than vitiates the increase of the far-IR radiation density of about a few 100.

\subsection{Diffusion of dust grains by turbulence}
Our observations indicate the presence of an evolved dust grain population towards the rim of the globule. Here, we investigate possible physical mechanisms that may be responsible for the transportation of dust grains from core to the rim. 

The observed linewidth from \citet{Benson98}, using C$_{3}$H$_{2}$, provides a value of 0.37 km/s in B207. The turbulent velocity dispersion, $\sigma$, is related to the line width $\Delta$v by 
\begin{equation}
\sigma^{2} = \Delta v^{2} - ln(2) \frac{8kT}{m},
\end{equation}
where m is the mass of C$_{3}$H$_{2}$ molecule \citep{Myers83}. The calculated turbulent velocity dispersion is $\sigma = v_{gas} = 0.35$ km/s. The distance between the core centre and the rim is about 24,000 au and with the measured turbulent velocity it will require 0.33 Myr to transport dust grains from the core to the rim. For larger distances and lower turbulent speed, the time required will be greater but still less than $\sim$1.0 Myr, similar to the age for the globule. Hence, turbulence can be a potential mechanism for the transportation of evolved dust grains from the central dense core to the rim.

The unusual position of the class I protostar, IRAS 04016+2610 (L1489IRS), at the edge of the core suggest migration of the protostar from the core and a possible physical link between core and star, where the core feeds material onto the proto-stellar disc \citep{Brinch07}. The link between the core and IRAS 04016+2610 is observed in the Herschel-SPIRE images (Fig. \ref{fig:b207_500}). A recent study by \citet{Wong16} also suggests that jets and outflows from protostars may contribute to the transportation of evolved dust grains.

\section{Conclusions}
We present optical observations of B207 made with the Large Monolithic Imager (LMI) instrument attached to the Discovery Channel Telescope (DCT) in U, B, V, R, and I bands, together with NIR, MIR and FIR data from 2MASS, WISE, and the Herschel Observatory. Physical properties of B207 are very similar to those found in other Bok globules in our galaxy. The main conclusions of our study are:\\
1. The temperature of the core measured from SED fitting and the Abel inversion method is 11.1 and 9.4 K, respectively. The temperature estimated from the Abel inversion technique is 1.7 K lower than found by the SED technique. However, similar temperatures are obtained at the rim region with both methods.\\
2. The column density, N(H) measured towards the core is about 7.2$\times$ 10$^{22}$ cm$^{-2}$, causing an extinction of A$_{V}$ = 48 mag for R$_{V}$ = 5.5. The column density gradient towards the south-east is steeper than in the north-west along our selected strip region, indicating the presence of compressed gas in the south-east. The number density at the core, n$_{c}$, is 5.6$\times$10$^{5}$ cm$^{-3}$.\\
3. The estimated mass of the core is about 5.6 M$_{\odot}$ along our selected strip region. The total mass of B207 in the 12.5$\arcmin$ $\times$ 12.5$\arcmin$ field of view of the LMI is about 17 M$_{\odot}$.\\
4. The dust albedo in the cloud increases with wavelength from 0.54 in the U band to 0.84 in the I band, indicating grain sizes substantially in excess of those found in the ISM.\\
5. The measured intensities at near- to mid- to far-infrared wavelengths at the core and rim regions of B207 are in agreement with the CMM+AMMI and CMM+AMM dust models, indicating the presence of large grains distributed throughout the entire globule.\\
6. Although MIR photons from the protostar are able to penetrate to the core and are comparable in intensity to the ISRF, they do not add significantly to the heating of the core.\\
7. The jets and outflows from the Class I protostar and a possible link between the core and IRAS 04016+2610 with a turbulence time scale of 0.33 Myr suggest that evolved dust grains can be diffused from the core to the rim within the estimated age of the globule. However, the possibility of in-situ formation of large dust grains in the rim region cannot be excluded.\\
\\
\textit{Acknowledgements}
We thank the anonymous referee for his or her thorough review and highly appreciate the comments and suggestions, which significantly contributed to improving the quality of the publication. We would like to thank A. Jones and N. Ysard for sharing the wavelength dependent plots of dust albedos and intensities. DSJ would like to thank the Physics and Astronomy REU programme of the University of Toledo, supported by the NSF grant No. 1262810. These results made use of Lowell Observatory's 4.3m Discovery Channel Telescope. Lowell operates the DCT in partnership with Boston University, Northern Arizona University, the University of Maryland, and the University of Toledo. Partial support of the DCT was provided by Discovery Communications. LMI was built by Lowell Observatory using funds from the National Science Foundation (AST-1005313). This publication makes use of data products from the Two Micron All Sky Survey, which is a joint project of the University of Massachusetts and the Infrared Processing and Analysis center and the California Institute of Technology, funded by the National Aeronautics and Space Administration and the National Science Foundation. This publication also makes use of data products from the Wide-field Infrared Survey Explorer, which is a joint project of the University of California, Los Angeles, and the Jet Propulsion Laboratory/California Institute of Technology, funded by the National Aeronautics and Space Administration. This paper uses data from \emph{Herschel}'s photometer PACS and SPIRE. PACS has been developed by a consortium of institutes led by MPE (Germany) and including UVIE (Austria); KU Leuven, CSL, IMEC (Belgium); CEA, LAM (France); MPIA (Germany); INAF-IFSI/OAA/OAP/OAT/LENS,SISSA(Italy); IAC (Spain). This development has been supported by the funding agencies BMVIT (Austria), ESA-PRODEX (Belgium), CEA/CNES (France), DLR (Germany), ASI/INAF (Italy), and CICYT/MCYT (Spain). SPIRE has been developed by a consortium of institutes led by Cardiff University (UK) and including: Univ. Lethbridge (Canada); NAOC (China); CEA, LAM (France); IFSI, Univ. Padua (Italy); IAC (Spain); Stockholm Observatory (Sweden); Imperial College London, RAL, UCL-MSSL, UKATC, Univ. Sussex (UK); and Caltech, JPL, NHSC, Univ. Colorado (USA).

\bibliography{mybib2}
\bibliographystyle{aa}
\end{document}